\documentclass{amsart}
\usepackage[latin1]{inputenc}
\usepackage[swedish,english]{babel}
\usepackage{ifpdf}
\ifpdf
  \usepackage[pdftex]{graphicx}
  \usepackage{epstopdf}
\else
  \usepackage[dvips]{graphicx}
\fi
\usepackage[pdftitle={Dynamic autotuning of fast multipole methods},
  pdfauthor={Holm, Engblom, Goude, Holmgren},
  pdffitwindow=true,
  breaklinks=true,
  colorlinks=true,
  urlcolor=blue,
  linkcolor=red,
  citecolor=red,
  anchorcolor=red]{hyperref}
\usepackage{slashbox,pict2e}
\usepackage{algorithmic}
\usepackage[ruled,norelsize]{algorithm2e}
\usepackage[numbers,sort]{natbib}

\numberwithin{equation}{section}
\numberwithin{table}{section}
\numberwithin{figure}{section}

\theoremstyle{plain}

\theoremstyle{definition}

\theoremstyle{remark}

\newcommand{\nlevels}{N_{\mbox{{\tiny levels}}}}
\newcommand{\thetacriterion}{$\theta$-criterion}
\newcommand{\TOL}{\operatorname{TOL}}
\newcommand{\realdom}{\mathbf{R}}
\newcommand{\Ordo}[1]{\mathcal{O}\left(#1\right)}
\newcommand{\PtoP}{\mbox{{\tiny P2P}}}
\newcommand{\MtoL}{\mbox{{\tiny M2L}}}
\newcommand{\MtoM}{\mbox{{\tiny M2M}}}
\newcommand{\LtoL}{\mbox{{\tiny L2L}}}
\newcommand{\PtoM}{\mbox{{\tiny P2M}}}
\newcommand{\LtoP}{\mbox{{\tiny L2P}}}

\begin{document}

\title[Dynamic autotuning of fast multipole methods]{Dynamic
  autotuning of adaptive fast multipole methods on hybrid multicore
  CPU \& GPU systems}

\author[M. Holm]{Marcus Holm}

\thanks{Corresponding author: S. Engblom, telephone +46-18-471 27 54, fax
  +46-18-51 19 25.}

\author[S. Engblom]{Stefan Engblom}

\author[A. Goude]{Anders Goude}

\author[S. Holmgren]{Sverker Holmgren}

\address[M. Holm, S. Engblom, and S. Holmgren]{Division of Scientific
  Computing, Department of Information Technology, Uppsala University,
  SE-751 05 Uppsala, Sweden.}

\email{marcus.holm, stefan.engblom, sverker.holmgren@it.uu.se}

\address[A. Goude]{Division of Electricity, Department of Engineering
  Sciences, Uppsala University, SE-751 21 Uppsala, Sweden.}

\email{anders.goude@angstrom.uu.se}


%65-XX Numerical analysis
%65Yxx Computer aspects of numerical algorithms
%65Y05 Parallel computation
%65Y10 Algorithms for specific classes of architectures
%65Y15 Packaged methods
%65Y20 Complexity and performance of numerical algorithms


\date{March 17, 2014}

\begin{abstract}
  We discuss an implementation of adaptive fast multipole methods
  targeting hybrid multicore CPU- and GPU-systems. From previous
  experiences with the computational profile of our version of the
  fast multipole algorithm, suitable parts are off-loaded to the GPU,
  while the remaining parts are threaded and executed concurrently by
  the CPU. The parameters defining the algorithm affects the
  performance and by measuring this effect we are able to dynamically
  balance the algorithm towards optimal performance. Our setup uses
  the dynamic nature of the computations and is therefore of general
  character.
\end{abstract}

\selectlanguage{english}

\maketitle

%**************************************************************************

\section{Introduction}

The $N$-body simulation is an ubiquitous problem in computational
science, arising in many different application areas and attracting a
lot of interest from developers of numerical algorithms and software
for many years. For the computation of interaction forces, the naive
all-pairs algorithm scales as $\mathcal{O}(N^2)$, and approximative
algorithms have been developed with better asymptotic computational
complexity. For large $N$-body problems requiring accurately
determined forces it is known that the Fast Multipole Method (FMM)
\cite{AFMM, FMM} is more efficient (in terms of the number of
arithmetic operations required) than both the all-pairs algorithm and
more classic tree-based schemes like the Barnes-Hut algorithm
\cite{Barnes_Hut,Dehnen02}.

About a decade ago, physical constraints in chip design spawned a
paradigm shift in computer architecture. Increased performance is now
realized mainly by increased thread parallelism rather than increased
clock speed. Also, the instruction complexity and the
memory/communication bandwidth is increasing at a significantly slower
rate than the computational performance. This implies that other
aspects than arithmetic complexity need to be assessed when designing
numerical algorithms, prominently parallelism and locality of data
access. In this context, FMM algorithms have the potential of becoming
increasingly important tools for CSE applications since they combine
optimal $\mathcal{O}(N)$ complexity with potential for large-scale
parallelism and large amounts of spatial and temporal data
locality. However, parallelizing and localizing the FMM computations
to make them suitable for modern computer hardware is a non-trivial
task, and has also received considerable attention lately
\cite{Chandramowlishwaran:2010b,Cruz2011,
  AFMMgpu,Jinshi2010,Yokota_FMM_manycore}.
% +++ Minisymposium SIAM CSE 2013

In this paper we present a recently developed variant of the FMM
\cite{thetanote, AFMMgpu} and show that it can be efficiently adapted
for parallelization on modern computer systems. We exploit the
heterogeneity inherent in the algorithm to compose a highly effective
yet flexible hybrid parallel FMM scheme which exploits both multiple
CPU threads and an accelerator. By using a dynamic autotuning
technique, our implementation can exploit systems with different
hardware characteristics and adapt to different problem settings
without the need for explicitly modifying a large set of computer
architecture- and problem-dependent parameters. The algorithm also
achieves good performance on dynamic problems that change
characteristics dramatically over time. We argue that the approach of
basing implementations on heterogeneous computer systems on a hybrid
approach using the inherent heterogeneity of \emph{the algorithm} can
be very fruitful for many other computational science kernels, apart
from the FMM.

In Section~\ref{sec:fmm} we summarize our version of the adaptive FMM
and also briefly discuss the computational complexity. The hybrid
parallelization is described in some detail, including complexity
estimates, in Section~\ref{sec:parallelization}. Our autotuning
approach is presented in Section~\ref{sec:autotuning}, where we design
several autotuning regulators in an incremental way. One important
aspect is that we want a black-box regulator not requiring explicit
complexity estimates. Also, we devise an autotuning scheme where the
extra work performed for tuning is limited by a parameter -
essentially the only parameter that is explicitly needed from the
user. In Section~\ref{sec:applications} we perform computational
experiments for our parallel implementation of a 2D FMM and show that
the suggested autotuning algorithm provides good performance for
problems from different classes, including those whose characteristics
change dynamically. A concluding discussion is found in
Section~\ref{sec:conclusions}.

%**************************************************************************

\section{Fast multipole methods}
\label{sec:fmm}

Since first presented in \cite{FMM,AFMM}, Fast Multipole
Methods (FMMs) have remained a crucial tool for fast evaluation of
pairwise interactions of the type
\begin{align}
  \label{eq:paireval}
  \Phi(x_{i}) &= \sum_{j = 1, j \not = i}^{N} G(x_{i},x_{j}),
  \quad x_{i} \in \realdom^{D}, \quad i = 1 \ldots N,
\end{align}
where $D \in \{2,3\}$ and where the kernel $G$ satisfies suitable
growth- and regularity assumptions \cite{thetanote}. Up to some
specified tolerance, the FMM algorithm produces a representation of
the field $\Phi(y)$ due to the $N$ \emph{sources} $\{x_{j}\}$ enclosed
in some finite domain (the terms \emph{potentials} and
\emph{particles} will also be used). Hence a slightly more general
viewpoint is that the FMM makes it possible to efficiently evaluate
\begin{align}
  \label{eq:poteval}
  \Phi(y_{i}) &= \sum_{j = 1, x_{j} \not = y_{i}}^{N} G(y_{i},x_{j}),
  \quad i = 1 \ldots M,
\end{align}
in which the effect of the sources $\{x_{j}\}$ is to be measured in a
set of \emph{evaluation points} $\{y_{i}\}$. In this section we
briefly describe the version of FMMs considered in this paper (see
\cite{thetanote,AFMMgpu} for earlier accounts).

\subsection{Well-separated sets}

FMMs are all based on the observation that the field experienced from
distant potentials can be approximated effectively. Generally, let a
collection of potentials be organized in two disjoint boxes with radii
$r_{1}$ and $r_{2}$, and let those boxes be separated at a distance
$d$. For quite general non-oscillating kernels $G$, one can then show
that the correct interpretation of `distant' is that \cite{thetanote},
\begin{align}
  \label{eq:thetacriterion}
  R+\theta r \le\theta d,
\end{align}
where $R = \max\left\{r_{1},r_{2}\right\}$, $r = \min\left\{
r_{1},r_{2}\right\}$, and $\theta \in (0,1)$ a certain parameter which
controls the accuracy. Eq.~\eqref{eq:thetacriterion} is the
\emph{\thetacriterion} and the two boxes are said to be
\emph{well-separated} whenever it applies. If this is so, then the
interactions can be compressed and handled as one single operation (or
\emph{shift}) between the boxes. If the criterion does not apply, then
the boxes are split into smaller boxes until they either are small
enough that \eqref{eq:thetacriterion} applies, or until they contain a
sufficiently small number of potentials that the interactions can be
computed directly.

\begin{figure}[htbp]
  \centering
  \includegraphics{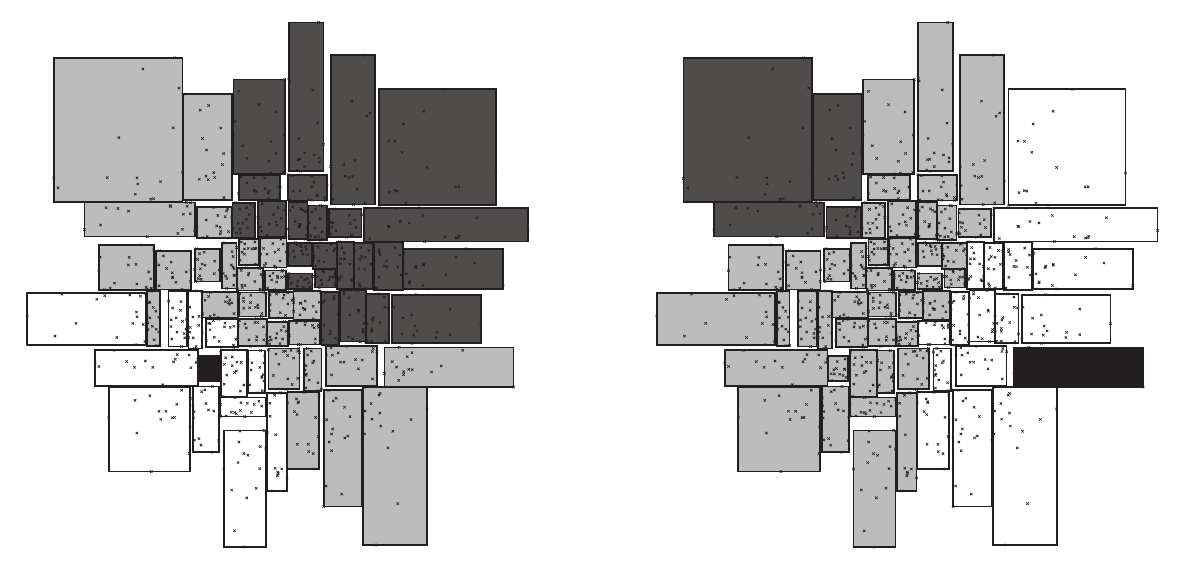}
  \caption{The different types of connections in an adaptive mesh. The
    target box is colored in black and \emph{strongly connected} boxes
    are white. The other boxes are \emph{well-separated} from the
    target box (they satisfy the
    \thetacriterion~\eqref{eq:thetacriterion}). The boxes in light
    gray are \emph{weakly coupled} and interact through M2L
    interactions at this level in the multipole tree, while the boxes
    in dark gray are \emph{decoupled} and has already been accounted
    for at a coarser level in the multipole tree.}
  \label{fig:FMM_mesh}
\end{figure}

This recursive way of iteratively dividing the source points makes a
tree-based approach natural, where each level in the \emph{multipole
  tree} contains a collection of boxes. Initially, all potentials are
understood to be organized into a single box at the 0th level in the
tree. Recursively one then splits the boxes into smaller boxes
(``children'') such that the number of points per box
decreases. Following the prescription in \cite{thetanote}, a pairwise
relation between all boxes at the same level in the tree is now
defined. Boxes are said to be either \emph{strongly} or \emph{weakly
  coupled}, or they are \emph{decoupled}. Firstly, a box is defined to
always be strongly connected to itself. Secondly, boxes obtained by
splitting strongly connected boxes are by default also strongly
connected. \emph{However}, if two such boxes happen to satisfy the
\thetacriterion, then they become weakly coupled. Finally, children of
weakly coupled boxes are defined to be decoupled. The result of this
way of handling the multipole mesh is visualized in
Figure~\ref{fig:FMM_mesh} where two examples of connectivity patterns
are displayed.

At all levels in the multipole tree, the boxes have an \emph{outgoing}
and an \emph{ingoing} expansion. The former, also referred to as a
\emph{multipole expansion}, represents the effect of potentials
enclosed in the box and is accurate in boxes which are well-separated
from the current box. The latter is instead a \emph{local} expansion
and captures the effect of distant well-separated boxes within the
current box.

% +++
%\begin{align}
%  \label{eq:multipole}
%  M(z) &= a_{0} \log(z-z_{0})+\sum_{j = 1}^{p} \frac{a_{j}}{(z-z_{0})^{j}},
%\end{align}
%\begin{align}
%  \label{eq:local}
%  L(z) &= \sum_{j = 0}^{p} b_{j} (z-z_{0})^{j},
%\end{align}

\begin{figure}[htbp]
  \centering
  \includegraphics{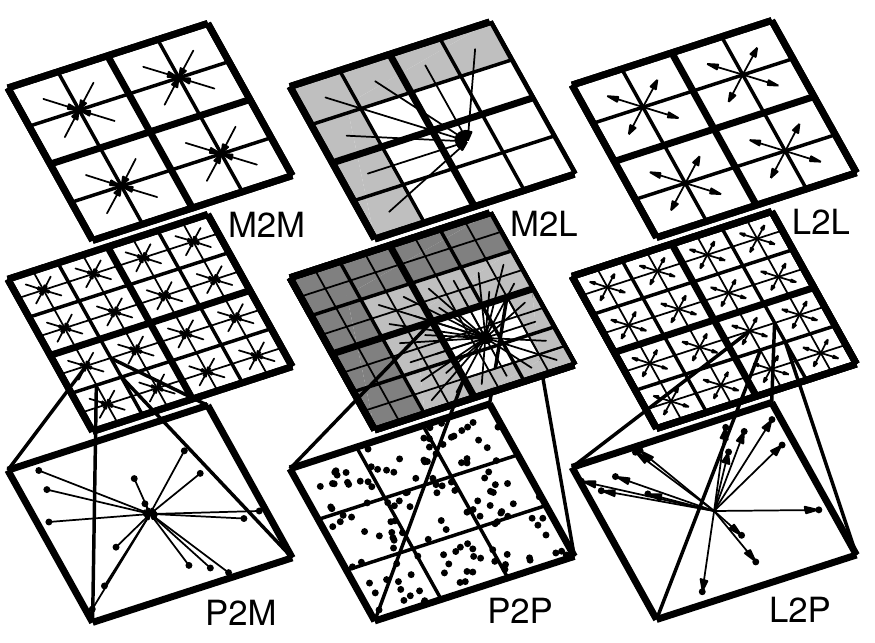}
  \caption{A schematic view of the FMM algorithm; see text for further
    details. \textit{Left:} Initialization using Particle-to-Multipole
    (P2M) shifts, and in the upward phase, Multipole-to-Multipole
    (M2M) shifts. \textit{Middle and right:} (downward phase)
    Multipole-to-Local (M2L), followed by Local-to-Local (L2L)
    shifts. \textit{Bottom:} The direct Particle-to-Particle (P2P)
    interaction and the Local-to-Particle (L2P) provides for the final
    evaluation of the potential field.}
  \label{fig:FMM_diag}
\end{figure}

The computational part of the FMM algorithm is depicted schematically
in Figure \ref{fig:FMM_diag}. Characteristically, one proceeds in an
\emph{upward} and a \emph{downward} phase. In the first phase, the
\emph{multipole-to-multipole} (M2M) operation propagates outgoing
expansion upwards in the tree. In the second phase, the
\emph{multipole-to-local} (M2L) and the subsequent
\emph{local-to-local} (L2L) operation translates and propagates this
field into local expansions downwards in the tree. Any remaining
potentials not accounted for through these operations are handled by
direct evaluation of \eqref{eq:paireval} at the finest level in the
tree.

\subsection{The balanced adaptive FMM}

When the FMM algorithm is described as above it makes no assumptions
on the precise way the boxes that make up the multipole mesh are to be
constructed. A standard implementation uses a tree data structure with
a locally adapted depth, resulting in a rather complex communication
pattern including several levels in the trees. An alternative
formulation is presented in \cite{thetanote}, where a \emph{balanced}
tree (or a \emph{pyramid} data structure) is used instead of a general
tree.  This introduces additional structure in the algorithm, making
parallelization easier and avoids communications across several levels
in the multipole tree. In order to balance the cost of the direct
evaluation at the finest level, splitting the boxes close to the
median value ensures an almost equal number of points in each box. For
more details, consult the distributed source code directly (see
Section~\ref{subsec:reproduce}). We refer to this FMM version as the
\emph{balanced} FMM algorithm. This scheme has previously been
implemented for 2D FMMs on a GPU \cite{AFMMgpu}, showing good
scalability and indeed indicating a potential for efficient
implementation also on other parallel architectures.

\subsection{Complexity}
\label{subsec:Complexity}

Most experience suggests that the practical complexity of the FMM
algorithm is $\Ordo{N}$ \cite{nbodycompare}, but it has been pointed
out that certain special distribution of points may imply a quadratic
complexity \cite{FMMcomplexitynote}. To get some feeling for the
computational complexity of the balanced FMM algorithm we first
consider a 2D setting and a collection of $N$ uniformly distributed
particles in the unit square. Let the FMM algorithm be parametrized by
the parameter pair $(\nlevels,\theta)$. Then the dominating
computational work is done at the finest level in the multipole tree
and amounts to \textit{(i)} the M2L shifts, and \textit{(ii)} the
direct P2P interactions.

Since the tree is a balanced quad-tree, the number of boxes at the
finest level is explicitly given by $N_{f} = 4^{\nlevels-1}$. Our
simplifying assumption of a uniform particle distribution implies that
the boxes are approximately uniform in size. Hence the radii are given
by
\begin{align}
  r &\sim 1/\sqrt{2N_{f}}, \\
  \intertext{and also}
  r_{\mbox{{\tiny parent}}} &\sim 2r,
\end{align}
where $r_{\mbox{{\tiny parent}}}$ is the radius at the second finest
level. Similarly, the average area of a box at the finest level is
given by $a := 1/N_{f}$ and the number of source points per box by
$n_{p} := N/N_{f}$. Thanks to the type of adaptivity used we note that
the latter estimate is independent of the details of the distribution
of points.

Consider first the arithmetic cost of the direct interaction
(P2P). Since we have assumed the radii to be approximately uniformly
distributed, $R \sim r$ in \eqref{eq:thetacriterion}, and hence the
boundary for direct interaction is found at $d \sim (1+\theta)/\theta
\times r$. With an area density of potentials $\rho := n_{p}/a$ we
get, since in each of $N_{f}$ boxes, $n_{p}$ points are to interact
with all points in a circle of radius about $d$, that the total
complexity can be estimated by
\begin{align}
  \nonumber
  C_{\PtoP} &\sim N_{f} \times \pi d^{2} \rho \times n_{p} \\
  \label{eq:Cp2p}
  &\sim \frac{N^{2}}{2N_{f}} \times \pi [(1+\theta)/\theta]^{2}.
\end{align}

Next we take the cost of the M2L-shift into account. Using $p$ terms
in both the outgoing and the ingoing expansions, we have that the
M2L-interaction is a linear mapping between $p$ coefficients, and
hence has complexity $p^{2}$. This mapping is performed in $N_{f}$
boxes provided that the \thetacriterion\ is true at the finest level,
but false at the second finest level. This can be written as
\begin{align}
  \nonumber
  C_{\MtoL} &\sim N_{f} \times \pi 
  (d_{\mbox{{\tiny parent}}}^{2}-d^{2})/a \times p^{2} \\
  \label{eq:Cm2l}
  &\sim \frac{3N_{f}}{2}p^{2} \times \pi [(1+\theta)/\theta]^{2},
\end{align}
where by the same argument as before $d_{\mbox{{\tiny parent}}} \sim
(1+\theta)/\theta \times r_{\mbox{{\tiny parent}}}$. Note that the
\emph{total} cost of M2L at all levels forms a geometric series in
terms of the work done at the finest level. Hence the total complexity
can be estimated to be $(1+1/4+1/16+...)  \times C_{\MtoL} \le 4/3
\times C_{\MtoL}$.

For completeness, let us also briefly discuss the other operations in
the FMM-algorithm. Thanks to the pyramid data-structure we readily see
that regardless of the distributions of points we always have for the
\emph{total} costs that
\begin{align}
  C_{\MtoM} &\sim C_{\LtoL} = (1+1/4+1/16+...) \times N_{f} \times p^{2} \le 
  4/3 \times N_{f} p^{2}, \\
  \intertext{and also trivially,}
  C_{\PtoM} &\sim C_{\LtoP} \sim N p.
\end{align}
With a specified relative tolerance $\TOL$, we have that $p \sim
\log\TOL/\log\theta$ (see \cite{thetanote}), so that by choosing
$N_{f} \propto N$, the total complexity can be expected to be bounded
by a constant times $\theta^{-2} \log^{-2}\theta \cdot N\log^{2}
\TOL$.

For a 3D FMM, the estimates corresponding to
\eqref{eq:Cp2p}--\eqref{eq:Cm2l} are
\begin{align}
  C_{\PtoP} &\sim \frac{3^{1/2}N^{2}}{2N_{f}} \times \pi
  [(1+\theta)/\theta]^{3}, \\
  C_{\MtoL} &\sim \frac{7 \cdot 3^{1/2} N_{f}}{2}p^{4} \times \pi
  [(1+\theta)/\theta]^{3},
\end{align}
where the factor $p^{4}$ can be improved to $p^{3}$, or even $p^{2}$,
depending on the implementation \cite{new_AFMM}. The total complexity
is in any case now bounded by a constant times $\theta^{-3}
\log^{-4}\theta \cdot N\log^{4} \TOL$.

%**************************************************************************

\section{Parallelization}
\label{sec:parallelization}

In this section, we describe an efficient and robust parallel FMM
algorithm for a heterogeneous computational node with several CPU
threads and a hardware accelerator in the form of a GPU. This is the
standard architecture for laptop and desktop computers and also the
standard building block in larger, more specialized computers aimed at
solving large-scale computational science and engineering problems. An
efficient single-node algorithm is clearly an essential building-block
when devising an implementation on multi-node distributed computer
systems. In Section~\ref{sec:applications} we present results from an
implementation of our autotuned parallel algorithm for a 2D FMM. As we
will clarify later, the parallelization scheme is applicable also to
3D settings, using the same basic data structures and autotuning
techniques.

\subsection{The parallel FMM algorithm for a hybrid node architecture}
\label{subsec:hybridfmm}

Devising algorithms for efficient utilization of a heterogeneous
computer architecture is a challenging task.  New hardware is
continuously developed and released, and there is a need for
algorithms than can easily be made efficient also on nodes with new
processors and accelerators. The algorithms should automatically adapt
to the specifics of new hardware, without the need for a user to
provide computer system-dependent parameters. In the same way, the
algorithm should be robust to changes in the problem setting,
relieving the user also of the task of modifying problem-dependent
parameters. Combined with the autotuning scheme discussed in
Section~\ref{sec:autotuning}, the parallel FMM algorithm presented
below fulfills these criteria. The algorithm presented can be very
useful for performing practical FMM computations in applications.

The balanced FMM algorithm has previously been implemented in a
GPU-only code \cite{AFMMgpu}. Here, good speedup is achieved compared
to a well-optimized single-core CPU implementation. However, not every
stage of the algorithm is equally well-suited for execution on the
GPU, and some stages are also independent and can be executed in
parallel. These are also the main observations forming the basis for
our new parallel FMM algorithm for hybrid architectures. First we note
that the most demanding stages in the FMM, the downward pass and the
direct near-field evaluations, are independent of each other. This
means that we can offload the near-field evaluations to the
accelerator while the downward pass is simultaneously completed on the
CPU. This pre-defined scheduling of the work is motivated by the fact
that small all-pairs $N$-body problems can be solved extremely
efficiently on accelerators, see e.g. \cite{GPUgems}, while the
hierarchical FMM operations can be considered to be better suited for
a general-purpose processor, at least in the sense of coding
complexity \cite{AFMMgpu}.

The pyramid data-structure used in the balanced FMM allows us to
easily parallelize the downward pass in the algorithm using a
task-based parallelization model. Here, a main task starts to work at
the root of the tree and is allowed to continue in a breadth-first
fashion until it reaches a certain predefined level. The main task
then creates one subtask per parent node at this level. The number of
potential tasks grows rapidly after each level, and the initial serial
work can in general be ignored for an implementation on a
multithreaded CPU. Because of the connectivity guaranteed by our
partitioning scheme, each task created in this way is independent of
the other tasks and completes the downward pass for one branch of the
tree. In this way, it is possible to launch a number of subtasks that
is appropriate for the CPU and create work units with suitable
granularity and memory footprint to ensure good load balance and cache
utilization.

The original formulation of the FMM algorithm explicitly calculates
$G(x_i,x_j)$ and $G(x_j,x_i)$ simultaneously, halving the number of
interactions to be computed. However, this symmetry introduces a
dependency in the downward pass that requires task synchronization on
every level. In a task-based parallel framework it is possible to
implement this with a minimum of communication, but experiments using
the implementation described below showed that this still results in
poor scaling. Hence, in the current implementation, both $G(x_i,x_j)$
and $G(x_j,x_i)$ are calculated explicitly, increasing the number of
arithmetic operations but removing the need for synchronization at all
FMM tree levels.

With the two heaviest components of the FMM algorithm being executed
simultaneously by the CPU and the accelerator, the work for
partitioning the particles and initializing the finest level of the
tree (P2M) becomes significant. Here, particle partitioning is easily
parallelizable in the same way as the downward pass while the P2M
step is embarrassingly parallel and can be parallelized accordingly.

In the near-field evaluation (P2P), the contribution from all boxes
within the near-field of a box should be calculated at all evaluation
points of the box. Similar to the M2L translations, the number of
boxes in the near-field varies due to the adaptivity. On a
multithreaded CPU, the parallelization is trivial. On a GPU-like
accelerator, parallelization is also easy but more work is needed to
ensure locality and efficient use of the GPU memory.

\begin{figure}
  \includegraphics[width=0.8\columnwidth]{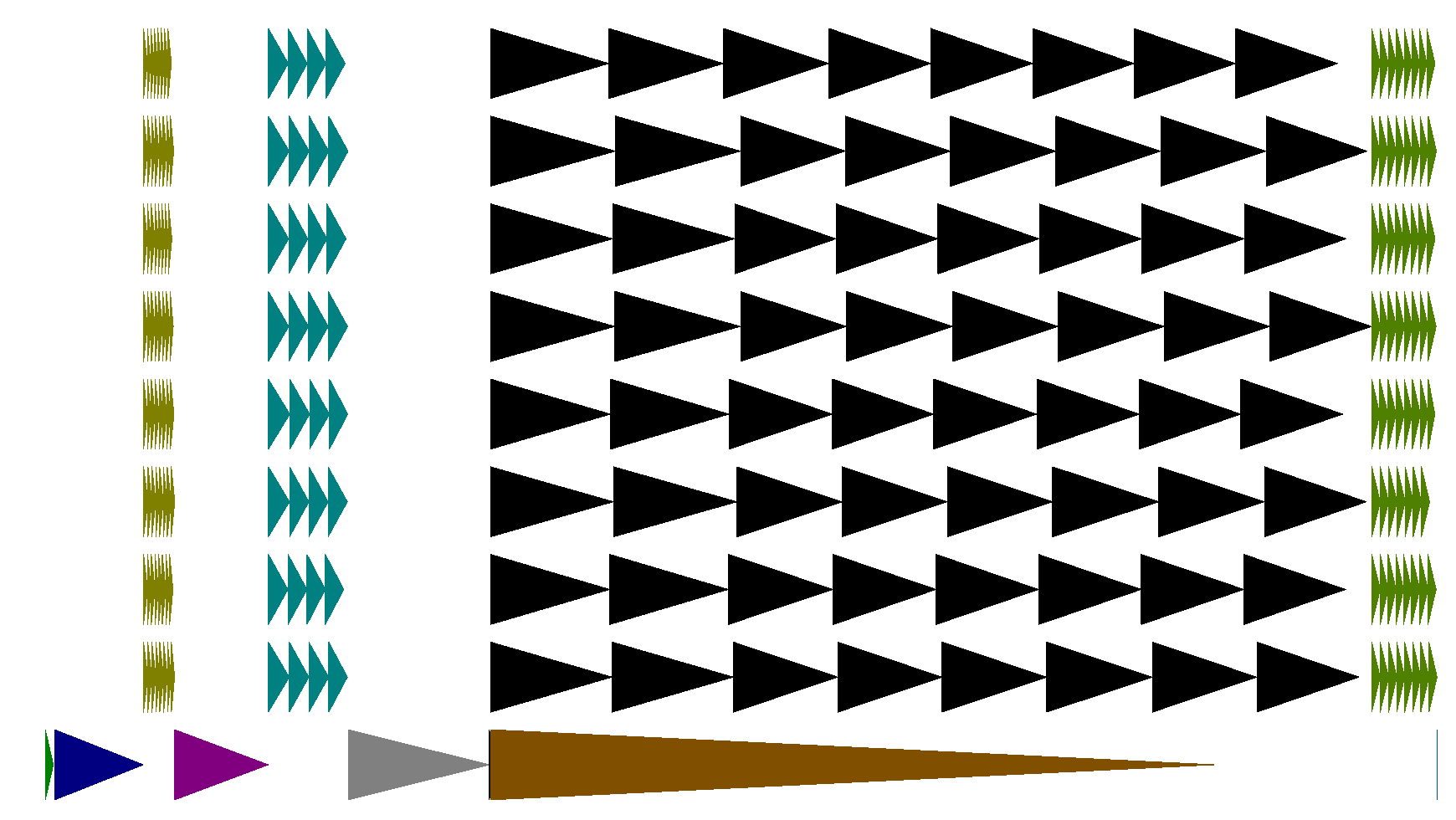}
  \caption{SuperGlue Execution Visualizator display for a typical run
    with eight worker threads and a GPU. Triangles correspond to
    tasks, and are executed from left to right. Each row represents
    one thread, starting with the master thread at the bottom. The
    parallel sections correspond to the initial topological phase
    (light green), the P2M- (cyan), the M2L- (black), and the
    L2P-shifts (green), respectively. The long brown arrow represents
    the main thread waiting while the P2P step executes on the
    GPU. \label{fig:sgvis}}
\end{figure}

In \cite{Chandramowlishwaran:2010b}, an efficient FMM algorithm for a
multithreaded CPU is described and implemented. Compared to the work
presented there, we have chosen a slightly different parallelization
strategy.  In \cite{Chandramowlishwaran:2010b}, the tree construction
is left serial, a choice motivated by the fact that in some
applications it may be possible to amortize this cost. We chose to
parallelize the particle partitioning phase of tree construction,
partly because our balanced FMM algorithm involves slightly more work
for this stage, but also because there are many applications in which
tree construction is done for every (or almost every) force
computation and amortizing is not efficient. We also made the choice
to leave the upward pass serial since this operation is less demanding
in our implementation than in the FMM variant in
\cite{Chandramowlishwaran:2010b} and since we do not expect this step
to become a significant performance bottleneck on hardware available
in the foreseeable future (see the length of the gray arrow in
Figure~\ref{fig:sgvis}).

\subsection{Implementation on a hybrid architecture}
\label{subsec:hybridimplementation}

Today, there is a plethora of different programming tools for
heterogeneous computer systems, still without a clear standard
emerging. To handle this situation, a clear separation between
algorithm and implementation is needed and algorithms should be easy
to implement using different programming tools.  However, to be able
to perform numerical experiments and performance tests, it is clear
that a choice of programming models must be made.  The current
parallel implementation of our FMM algorithm uses a task-based
parallelization approach implemented using the SuperGlue library
\cite{Tillenius10}. In task-based programming, a number of worker
threads is created at program initialization and execute tasks that
are submitted to a scheduler dynamically. There are other task-based
programming libraries that are designed with hybrid architectures in
mind, for example StarPU \cite{AugThiNamWac11CCPE}. Here, a black-box
approach is taken, where the scheduler is used to execute a given set
of tasks in an efficient way, without taking any specifics of the
underlying algorithm into account. StarPU is able to choose among
hardware resources and select where to schedule tasks so the total
runtime will be small. For example, if a relatively fast GPU has a
long queue of tasks already scheduled, StarPU may decide to give a
task to a relatively slow CPU thread instead. Hence, a tool like
StarPU works with a static algorithm that produces a set of tasks that
are dynamically scheduled and executed. In contrast, our approach is
to tune the \emph{algorithm} itself dynamically, producing a mixture
of tasks that can be efficiently executed on a heterogeneous machine
essentially without requiring a unified task library.

% +++
% By working on the algorithm level, we gain the benefits of
% heterogeneous architectures without paying the costs of writing
% multiple kernel implementations.

Our current FMM implementation assumes the use of a GPU accelerator,
and the accelerator code is written using Nvidia Cuda 2.0
\cite{cudaguide}. As indicated above, we realize that for future
implementations, another task-based programming model and/or another
other programming tool for the accelerator should possibly be used.

\subsection{Parallel efficiency and attainable performance}

Before we describe our methods for autotuning, we will show that our
FMM implementation is efficient and achieves satisfactory hardware
utilization. The tests described below were performed using the
hardware configuration described in Section~\ref{sec:applications}.

\begin{figure}
  \includegraphics{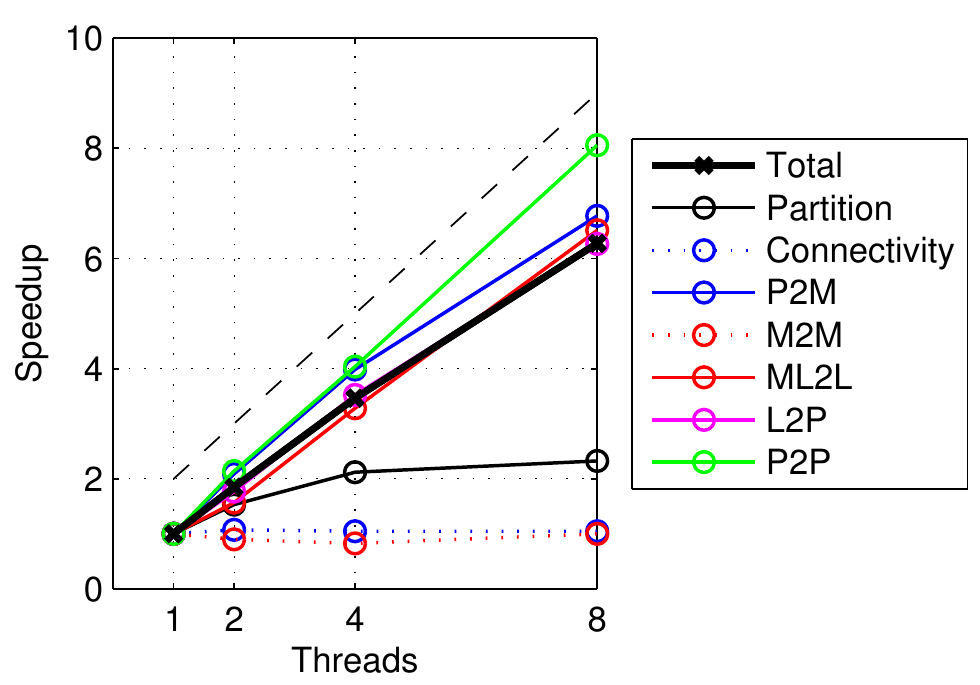}
  \caption{Strong scaling using only CPU, after tuning. Phases represented by
    dotted lines were not parallelized.}
  \label{figs:cpuspeedup}
  % +++ figure is problematic in black/white (...)
\end{figure}

The work in \cite{Chandramowlishwaran:2010b} provides a good reference
for comparison when using only a multithreaded CPU. In
\cite{Chandramowlishwaran:2010b}, good speedup for the P2P phase on up
to 8 threads on an Intel Nehalem is achieved. However, the other
computationally heavy phases show speedups between about 3.2x (for
M2L) and 5x (for M2M and P2M). Figure~\ref{figs:cpuspeedup} shows the
speedup for our implementation for up to 8 threads and for a problem
with a million uniformly distributed particles. It is clear that the
speedup is perfect for the P2P phase, and about 6--7x for most other
phases. The partitioning step of tree construction suffers from a
non-optimal data access pattern and is bandwidth-bound, resulting in a
maximum speedup of 2.3x. The total speedup on 8 threads for the full
multithreaded algorithm is 6.26x.

\begin{figure}
  \includegraphics{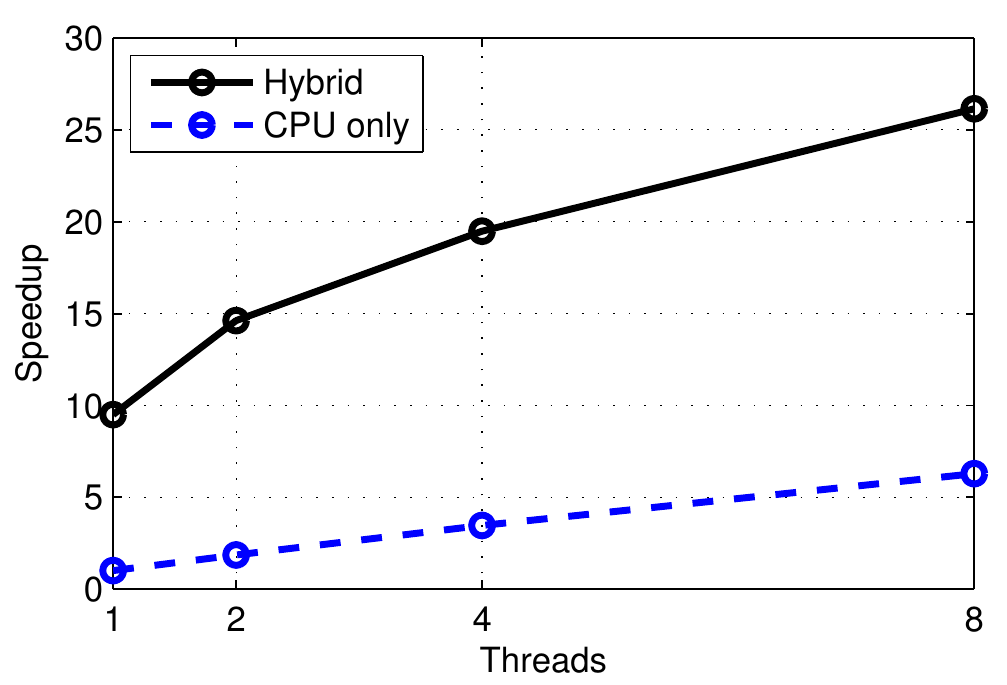}
  \caption{Comparison of CPU/GPU-hybrid and CPU-only speedup versus 1
    CPU-only thread.}
  \label{figs:gpucpuspeedup}
\end{figure}

In \cite{AFMMgpu} it was shown that the P2P phase of our algorithm can
run up to 12x faster on an Nvidia Tesla C2075 GPU compared to a single
Intel Xeon W3680 core, and the code as a whole ran about 10x faster on
the GPU.  A straight comparison with those results is not appropriate
because the codes are optimized differently (for example, to reduce
the memory footprint, we forego a reordering that enables the use of
SSE instructions and improves cache performance, which was implemented
in \cite{AFMMgpu}). Figure~\ref{figs:gpucpuspeedup} shows that using a
single CPU core plus the GPU yields a runtime that is 9.5x shorter
than with just one CPU core, which is quite encouraging considering
that only the P2P step is offloaded to the GPU. The performance using
multiple threads and the GPU is here 26x faster than the
single-threaded CPU-only code.

In Figure~\ref{figs:gpucpuspeedup}, we also see that using the full
CPU plus the GPU gives a performance that is 4.2x faster than using
only the CPU, so adding one GPU to the system is equivalent to adding
at least three CPU sockets in this implementation, assuming perfect
speedup on the CPU. Using the GPU instead of additional CPUs, we avoid
hitting the memory bandwidth limit and benefit from the higher
floating point capabilities of the GPU.

The performance measurements described above were made using the
harmonic potential $G(x_{i},x_{j}) = -m_{j}/(x_{i}-x_{j})$, where
$x_i$ and $x_j$ are complex numbers. We also implemented the
computationally more expensive logarithmic potential, used for example
in computer graphics when plotting isopotentials. The additional
arithmetic intensity of the logarithmic kernel improves the efficiency
of our results because of better amortization of data transfer costs
and kernel launch times on the GPU, and also for better cache
performance on the CPU. Using the logarithmic potential, the CPU+GPU
performance is 6x faster than using just the CPU, and the full system
performance is 40x faster than our single-threaded CPU-only
performance.

%**************************************************************************

\section{Autotuning}
\label{sec:autotuning}

Our autotuning scheme continuously measures the performance of
relevant FMM subtasks and makes autotuning decisions for two
performance-critical parameters in the FMM algorithm. Here, the only
crucial assumption is that the FMM is used in a time-marching or
iterative context. For most $N$-body problems, this is indeed the
case. No parameters estimating the complexity of subtasks or computer
system specifics need to be specified by the user.

The algorithmic complexity of the main program phases, as determined
by $\theta$ and $\nlevels$, was described in
Section~\ref{subsec:Complexity}. For a given relative error tolerance
TOL and $\theta$, we use the result from \cite{thetanote} that $p
\propto \log\TOL/\log\theta$ to choose an expansion degree $p$ to
satisfy the required tolerance. This lets us use $\theta$ as a
performance-tuning parameter while maintaining an error tolerance
appropriate to the problem. Table \ref{tab:ptoltheta} gives a sense of
the range of $p$.

\begin{table}[htdp]
  \begin{center}
    \begin{tabular}{crrrrr}
     \hline
     \backslashbox{~~~TOL}{$\theta$} & 0.35 & 0.4 & 0.5 & 0.6 & 0.65 \\
     \hline
     $10^{-6}$ & 11 & 13 & 17 & 24 & 28 \\
     $10^{-7}$ & 14 & 16 & 21 & 28 & 34 \\
     $10^{-8}$ & 16 & 18 & 24 & 33 & 39 \\
     \hline
   \end{tabular}
  \end{center}
  \caption{Expansion degree $p$ for sample values of TOL and $\theta$
    (harmonic potential).}
  \label{tab:ptoltheta}
\end{table}

Roughly speaking, the CPU-part that is executed concurrently with the
GPU-part is dominated by the downward pass, the runtime of which
decreases as the runtime of the GPU-part increases. This opens up for
an \textit{``Extremum Control''}-approach
\cite[Chap.~13.3]{AdaptiveControl} where these two parameters are
varied dynamically to remain close to the optimal choice. Clearly,
controlling $\theta$ and $\nlevels$ also controls the load-balance
between the CPU and GPU. Intuitively, this is desirable because we can
ensure maximum utilization of the available hardware.  As we will see,
however, maximum utilization does not necessarily correspond to
minimal runtime.

\subsection{Static tuning}
\label{subsec:statictuning}

Before we can describe our approach to autotuning in more detail, the
effects of parameter selection and particle distribution on the
performance of our FMM implementation need to be clarified. We divide
the performance-critical components of the code into three distinct
sections; the M2L phase, consisting of the downward pass through the
tree; the P2P phase, which consists of the direct evaluations in the
near field; and the $Q$ phase, which consists of the rest of the
program.  This division allows us to write the runtime of the hybrid
code $\tilde{C}_{\text{hybrid}}$ and the CPU-only code
$\tilde{C}_{\text{CPU}}$ for a given problem as functions of $\theta$
and $\nlevels$ in the following way:
\begin{align}
  \tilde{C}_{\text{hybrid}}(\theta,\nlevels) &= 
  \max( \tilde{C}_{\MtoL}(\theta,\nlevels),\tilde{C}_{\PtoP}(\theta,\nlevels)) + 
  \tilde{C}_{Q}(\theta,\nlevels), \\
  \intertext{and}
  \tilde{C}_{\text{CPU}}(\theta,\nlevels) &=
  \tilde{C}_{\MtoL}(\theta,\nlevels) + \tilde{C}_{\PtoP}(\theta,\nlevels) + 
  \tilde{C}_{\text{Q}}(\theta,\nlevels).
\end{align}

The runtime of the M2L phase is given by
$\tilde{C}_{\MtoL}(\theta,\nlevels) = A C_{\MtoL} + B$ for some
constants $A, B$ (see \eqref{eq:Cm2l}), and similarly for
$\tilde{C}_{\PtoP}$ (see \eqref{eq:Cp2p}). From the complexity
analysis in Section~\ref{subsec:Complexity} it is reasonably clear
that at least one minimum $(\theta,\nlevels)$ exists, but the precise
location depends on hardware, implementation, and problem specifics.

\begin{figure}
  \includegraphics{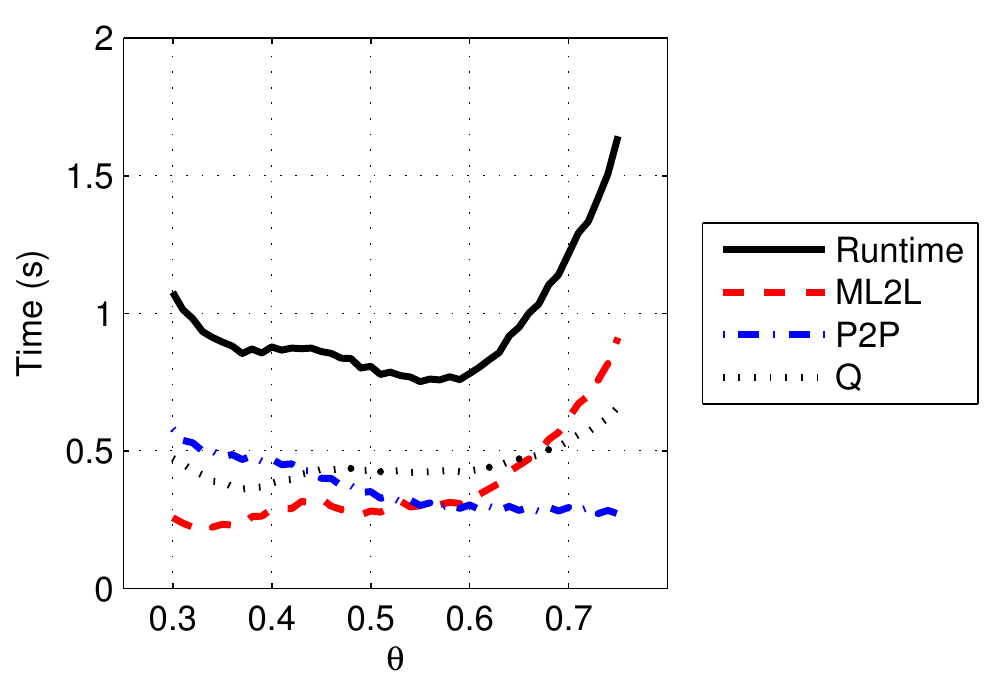}
  \caption{Runtime of hybrid code and the components M2L, P2P, and Q
    with varying $\theta$. Note the saw-tooth pattern in M2L, P2P, and
    total runtime, and also the ``central hump'' behavior in M2L and
    Q. Simulation performed with 8 CPU-threads, $N = 10^{6}$ and a
    uniform square particle distribution.}
  \label{fig:gpuvarytheta_uniform}
\end{figure}

When considering the computational work for the M2L, P2P, and Q
phases, it seems an attractive idea to tune
$\tilde{C}_{\text{hybrid}}$ such that the CPU and GPU parts are
balanced, $\tilde{C}_{\MtoL}(\theta,\nlevels) \approx
\tilde{C}_{\PtoP}(\theta,\nlevels)$, ignoring the small and nearly
constant runtime for the Q phase. However, once M2L is parallelized
and P2P is efficiently off-loaded to the GPU, Q becomes relatively
large and this strategy is not optimal (see Figure
\ref{fig:gpuvarytheta_uniform}). Therefore, we did not consider using
load-balance information to tune $\theta$, but only $\nlevels$. Being
run on an accelerator, P2P does not scale down with small problems
because of PCI latency and kernel startup time, so for small problems,
P2P is relatively constant. This means that optimal tuning for small
problems puts a seemingly disproportionate amount of work on the GPU,
because the time saved on performing Q on very small trees is greater
than the increase in time spent on P2P. This also means that the
performance gain of the GPU is smaller on small problems.

\begin{figure} 
  \includegraphics{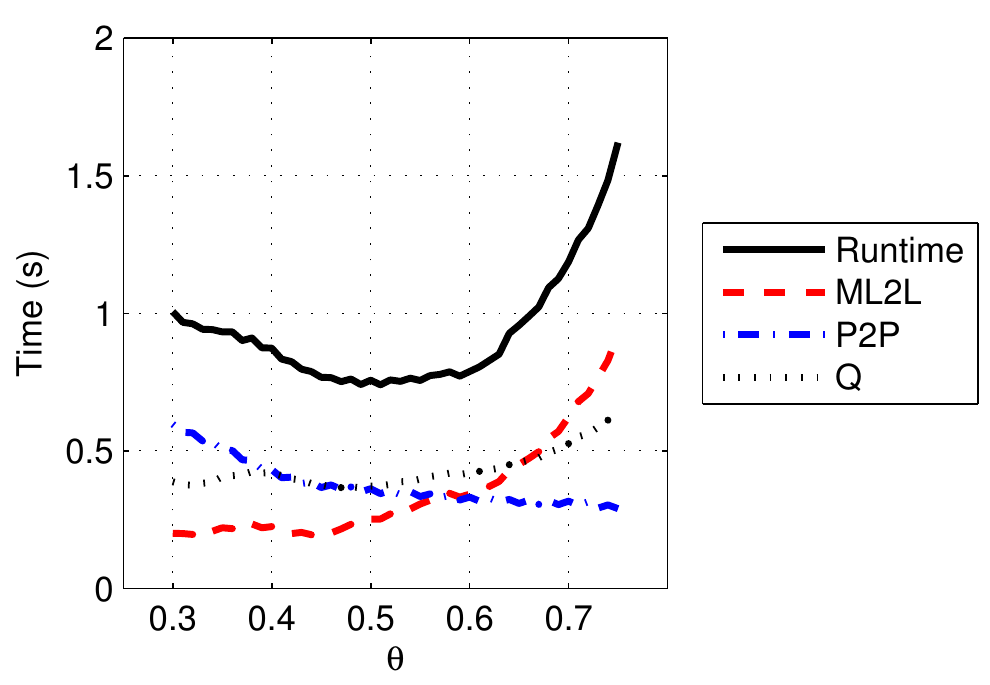}
  \caption{Runtime of line-like distribution. Note the ``shallower''
    and left-shifted minimum compared to the uniform
    distribution.}
  \label{fig:gpuvarytheta_linear}
\end{figure}

To study how particle distribution affects performance and tuning we
have run the same experiment as in Figure
\ref{fig:gpuvarytheta_uniform}, but this time with particles
distributed approximately along a line. This scenario can for example
be found in interface simulations and long Karman streets, as well as
the vortex instability simulation that we present in
Section~\ref{subsec:KHinstability}. The specific effects of changing
particle distributions depends on problem size, interaction potential,
and machine specification, but it is worth pointing out that tuning
for the wrong distribution can lead to poor performance. The
computations for the linear distribution (Figure
\ref{fig:gpuvarytheta_linear}) ran most efficiently with
$\theta=0.49$, while the uniform distribution ran most efficiently
with $\theta=0.55$. Running the uniform distribution with
$\theta=0.49$ would yield a performance penalty of 7\%.  In our
experience, the optimal value of $\theta$ may lie anywhere in the
range $[0.35, 0.65]$ and the potential performance penalty for values
in this range often exceeds 30\%. This motivates the use of dynamic
autotuning.

The reason why optimal performance and maximum utilization do not
always coincide is made clear in Figures
\ref{fig:gpuvarytheta_uniform} and \ref{fig:gpuvarytheta_linear}.
Maximum hardware utilization occurs when the M2L curve crosses the P2P
curve. Depending on where this balance occurs, the runtime cost of
lower utilization can be offset by savings in Q.

\subsection{Dynamic Autotuning}\label{sec:dynamic_autotuning}

The goal of autotuning is to achieve optimal performance while
maintaining \emph{generality} (applicability to many problems and
situations), \emph{robustness} (overcoming pathological conditions),
\emph{speed} (quickly finding the optimum), and \emph{efficiency}
(minimizing any additional computational work).

Generality is achieved by making as few assumptions as possible on the
system we try to control. The main assumption here is that each
iteration is a small incremental change compared to the previous one
such that the observed response in performance is due to tuning
attempts and not to the evolution of input data. Clearly, this
assumption is necessary in order to use the runtime per iteration to
evaluate the effectiveness of tuning parameters. A system which
drastically changes is problematic to control simply due to poor
estimation of the efficiency of the tuning steps.

A robust autotuner must be able to handle a number of potential
pitfalls. Here we identify and describe some of the difficulties that
we encountered in devising the autotuning for our FMM implementation.

\subsubsection{Noise}

Runtime measurements can vary for a multitude of reasons that are
unrelated to the actual tuning efficiency. Runtime variation that is
not attributable to changes in problem configuration or in tuning
parameters is called \emph{noise}. The presence of noise necessitates
taking repeated measurements. True repetition would involve freezing
the problem state for several calls to the FMM-routine, but this would
be prohibitively expensive. Instead, we assume that problem
configuration changes sufficiently slowly and use the minimum runtime
from a short sequence of iterations when making tuning decisions.

\subsubsection{Multiple local minima}

As Figure~\ref{fig:gpuvarytheta_uniform} shows, multiple local minima
in runtime may exist. We have seen two large near-optimal regions,
spaced widely apart in $\theta$, as well as a saw-tooth profile when
taking small steps in $\theta$.  An autotuner can get stuck at
suboptimal parameter values even when tuning a relatively static
problem. The reason for the two large near-optimal regions has been
difficult to analyze in detail, but is likely related to the way
$\theta$ is linked both to the number of expansion coefficients and
connectivity pattern for multipole boxes. The saw-tooth pattern is
similarly difficult to analyze conclusively, but we speculate that it
is caused by discretization effects on the cache or shared memory
efficiency of the M2L and P2P routines.  Below, we present how the
problem of multiple local minima can be solved using techniques from
global optimization.

\subsubsection{Discontinuous movement of the global minimum}

When the problem configuration shifts the relative efficiency of two
large near-optimal regions, the global optimum may move from the one
to the other. Capturing such an event exactly requires knowledge of
the performance of the entire range of $\theta$ at every iteration, so
we need a heuristic rule that is efficient and not to costly.

\subsubsection{Correlated controls}

If there are two (or more) tuning parameters, there exists the
possibility that each parameter cannot be tuned independently of the
other. For the tuning to improve, \emph{both} parameters must be
adjusted at once. However, both our static analysis and experiments
show that the optima in $\theta$ and $\nlevels$ are in practice almost
completely independent.

In order to avoid creating an excessively complicated and
application-specific scheme, we developed and evaluated a series of
autotuning techniques, each of which addresses a particular issue,
finishing with a robust autotuning system suitable for use in
applications (see Section~\ref{sec:applications}). The common design
principle is that each method periodically attempts a change in a
parameter (which we call a move), which is either accepted or rejected
depending on the performance in the following time-steps.

\subsubsection{AT1: Random walk} 

The simplest and most general autotuner performs a straightforward
biased random walk. For each parameter, moves are generated in
regularly spaced intervals as steps in a randomly selected direction.
A pseudocode description is provided in Algorithm~\ref{alg:at1}, where
$\texttt{time}_i$ and $p_i$ are the runtime and the parameter
configuration of the $i$th FMM call. $\texttt{thetastep}$ is the unit
length of a move in the $\theta$ parameter and is set to $0.01$ unless
otherwise specified and {\tt randbit} is either 0 or 1, chosen
randomly and with equal probability.

\begin{algorithm}
  \caption{AT1: Random walk \label{alg:at1}}
  \begin{algorithmic}
    \IF {$\texttt{time}_i > \texttt{time}_{i-1}$}
    \STATE return $p_{i+1} = p_{i-1}$ \COMMENT{Reject previous move}
    \ENDIF
    \IF {time to move in $\nlevels$}
    \STATE  return $p_{i+1} =[\theta, \nlevels + 
      (2 \cdot \texttt{randbit}-1)]$
    \ELSIF {time to move in $\theta$}
    \STATE  return $p_{i+1} = 
            [\theta+(2 \cdot \texttt{randbit}-1) \cdot 
              \texttt{thetastep},\nlevels]$
    \ELSE
    \STATE return $p_{i+1} = p_i$
    \ENDIF
  \end{algorithmic} 
\end{algorithm}

\subsubsection{AT2: Directed walk with varying step size}

A simple improvement is to remember the previous step. Moves are
generated in the same direction if the previous move was successful,
otherwise in the opposite direction. This method avoids moves in the
wrong direction, but it can get stuck in a local minimum such as the
ones observed in the static tuning experiments
(Section~\ref{subsec:statictuning}). In the pseudocode of
Algorithm~\ref{alg:at2}, $\texttt{move}$ represents a two dimensional
direction in parameter space.

\begin{figure}
  \includegraphics{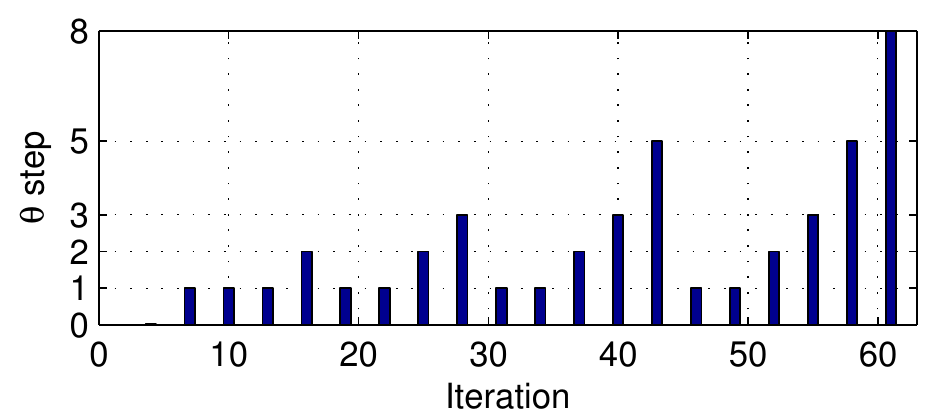}
  \caption{Progression of $\theta$ step length.\label{fig:wcycle}}
\end{figure}

In order to avoid the problem of getting stuck in a local minimum when
tuning $\theta$, we introduce a growing step size when an apparent
optimum is reached. For this to be effective, the step size must not
grow too slowly, but growing the step size too rapidly can cause the
algorithm to attempt big, large-grained, and expensive steps too
often. Furthermore, in order to track a moving optimum closely, most
step sizes must be small. We chose to use a sequence of steps which
cycles through the Fibonacci sequence (see
Figure~\ref{fig:wcycle}). In the description of
Algorithm~\ref{alg:at2}, the function fib($n$) returns the $n$th
Fibonacci number, and the counter $\texttt{fibcount}$ is initialized
to 1. Each time the end of a sequence is reached without accepting a
move, the counter is reset and the length of the sequence,
$\texttt{fiblength}$, is increased. For our applications, this
extension has not been necessary, as AT1 already successfully avoids
getting stuck in local minima, but it induces a negligible cost and is
potentially very useful.

\begin{algorithm}
  \caption{AT2: Directed walk with varying step size \label{alg:at2}}
  \begin{algorithmic}
    \IF {$\texttt{time}_i > \texttt{time}_{i-1}$}
    \IF {previous move was a $\theta$-move}
      \IF {$\texttt{fibcount} < \texttt{fiblength}$}
        \STATE set $\texttt{thetastep} = \text{fib}(\texttt{fibcount}++)$ 
      \ELSE
        \STATE set $\texttt{thetastep} = \text{fib}(\texttt{fibcount} = 1)$
        \STATE set $\texttt{fiblength++}$ \COMMENT{Grow sequence}
      \ENDIF
      \STATE set $\texttt{thetadir} = -\texttt{thetadir}$ \COMMENT{Reverse direction}
    \ELSE
      \STATE set $\texttt{Nldir} = -\texttt{Nldir}$ \COMMENT{Reverse direction}
    \ENDIF
    \STATE return $p_{i+1} = p_{i-1}$ \COMMENT{Reject previous move}
    \ENDIF
    \IF {time to move in $\nlevels$}
    \STATE set $\texttt{move} = [0, 1] \cdot \texttt{Nldir}$
    \ELSIF {time to move in $\theta$}  
    \STATE set $\texttt{move} =  [\texttt{thetastep}, 0]\cdot \texttt{thetadir}$
    \ELSE
    \STATE set $\texttt{move} = [0, 0]$
    \ENDIF   
    \STATE return $p_{i+1} = p_{i} + \texttt{move}$ 
  \end{algorithmic} 
\end{algorithm}

\subsubsection{AT3a: Loadbalance-aware directed walk with varying step size}

Intuition tells us that good hardware utilization yields good
performance. This information can be included in an autotuner that
generated moves in $\nlevels$ that always moves the loadbalance
towards zero. Note that for CPU-only runs, this method is equivalent
to AT2.

\begin{algorithm}
\caption{AT3a: Loadbalance-aware directed walk with
  varying step size \label{alg:at3a}}
\begin{algorithmic}
  \IF {$\texttt{time}_i > \texttt{time}_{i-1}$}    
  \IF {previous move was a $\theta$-move}
      \IF {$\texttt{fibcount} < \texttt{fiblength}$}
        \STATE set $\texttt{thetastep} = \text{fib}(\texttt{fibcount}++)$ 
      \ELSE
        \STATE set $\texttt{thetastep} = \text{fib}(\texttt{fibcount} = 1)$
      \ENDIF
      \STATE set $\texttt{thetadir} = -\texttt{thetadir}$     \COMMENT{Reverse direction}
  \ENDIF
  \STATE return $p_{i+1} = p_{i-1}$ \COMMENT{Reject previous move}
  \ENDIF
  \IF {time to move in $\nlevels$}
    \IF {CPU waits on GPU}
      \STATE  set $\texttt{move}= [0,1]$ \COMMENT{More work on the CPU}
    \ELSE
      \STATE  set $\texttt{move} =[0,-1]$ \COMMENT{More work on the GPU}
    \ENDIF
  \ELSIF {time to move in $\theta$}  
    \STATE set $\texttt{move} =  [\texttt{thetastep}, 0]\cdot \texttt{thetadir}$
  \ELSE
    \STATE set $\texttt{move} = [0, 0]$
  \ENDIF   
   \STATE return $p_{i+1} = p_{i} +  \texttt{move}$
\end{algorithmic} 
\end{algorithm}

Another approach for determining the correct step direction for
$\nlevels$ would be to use a scaling model that, based on the FMM's
algorithmics, would predict the change in workloads for the CPU and
the GPU.  We did not pursue this approach, first because it is too
application specific, and second because of the difficulty in
incorporating hardware- and problem-specific performance parameters
into a model based on algorithmics. For example, cache and memory bus
behavior can dramatically affect both the CPU and the GPU workload.

\subsubsection{AT3b: Directed walk with varying step size and cost estimation}

In this final version of autotuning, the user specifies
$\texttt{cap}$, a maximum cost for the autotuning of $\nlevels$.  When
a move is made that degrades performance, the next move in the same
direction is scheduled so that the expected runtime cost is less than the
specified maximum cost. The expected cost is based on  
the difference between the runtime of the rejected step and the runtime 
of the most recent accepted step.
Cost estimation has two benefits: it prevents
the cost for the autotuning to grow out of bounds, and since the test
frequency increases near switching points it captures these almost
exactly.

\begin{algorithm} 
  \caption{AT3b: Directed walk with varying step size and cost estimation \label{alg:at3b}}
 \begin{algorithmic}
  \IF {$\texttt{time}_i > \texttt{time}_{i-1}$}
  \IF {previous move was a $\theta$-move}
  \IF {$\texttt{fibcount} < \texttt{fiblength}$}
  \STATE set $\texttt{thetastep} = \text{fib}(\texttt{fibcount}++)$
  \ELSE
  \STATE set $\texttt{thetastep} = \text{fib}(\texttt{fibcount} = 1)$
  \ENDIF
  \STATE set $\texttt{thetadir} = -\texttt{thetadir}$
  \COMMENT{Reverse direction}
  \ELSE
  \STATE $\texttt{cost} = \texttt{time}_i - \texttt{time}_{i-1}$ 
  \IF {$\texttt{Nldir} > 0$}
  \STATE $\texttt{upcost} := \texttt{upcost} + \texttt{cost}$ 
  \COMMENT{cost of misstep}
  \STATE $\texttt{uptime} := (\texttt{upcost} + \texttt{cost})/\texttt{cap}-\texttt{basetime}$
    \COMMENT{time to next up move} 
  \STATE $\texttt{upinterval} := \texttt{uptime} \cdot i / \texttt{basetime}$  
    \COMMENT{\# iterations to next up move} 
  \ELSE
  \STATE \COMMENT{same logic for \emph{decreasing} $\nlevels$:}
  \STATE $\texttt{downcost} := \texttt{downcost} + \texttt{cost}$
  \STATE $\texttt{downtime} := (\texttt{downcost} + \texttt{cost})/\texttt{cap}-\texttt{basetime}$
  \STATE $\texttt{downinterval} := \texttt{downtime} \cdot i / \texttt{basetime}$
  \ENDIF
  \ENDIF    
  \STATE return $p_{i+1} = p_{i-1}$ \COMMENT{Reject previous move}
  \ELSE
  \STATE set $\texttt{basetime} = \texttt{basetime} + \texttt{time}_i$ 
  \ENDIF
  \IF {time to move in $\nlevels$}
  \STATE set $\texttt{move} = [0, \pm 1]$ as appropriate
  \ELSIF {time to move in $\theta$}  
  \STATE set $\texttt{move} =  [\texttt{thetastep}, 0] \cdot \texttt{thetadir} $
  \ELSE
  \STATE set $\texttt{move} = [0, 0]$
  \ENDIF   
  \STATE return $p_{i+1} = p_{i} + \texttt{move}$ 
   \end{algorithmic} 
   \end{algorithm}

%**************************************************************************

\section{Applications}
\label{sec:applications}

In this section, we illustrate the behavior of our autotuning
implementation by presenting results for three example applications
with different properties. The numerical experiments were performed on
single GPU-equipped nodes on a cluster. Each node is equipped with two
8-core AMD Opteron 6220 (Bulldozer) processors configured with 8
floating-point units in total and an Nvidia Tesla M2050 GPU. Timing
measurements reported are the full iteration times or complete
simulation times, as noted. Unless otherwise noted, the relative error
tolerance is set to $10^{-6}$ (see Table~\ref{tab:ptoltheta} for the
number of expansion coefficients).

\subsection{Vortex instability}
\label{subsec:KHinstability}

This problem involves a set of vortices, which are propagated with
the flow velocity according to
\begin{align}
  \frac{dx_{k}}{dt}=\frac{1}{2\pi i}\sum_{k=1}^{N}
  \frac{\Gamma_{i}}{\overline{x}-\overline{x}_{k}}
  g_{\delta}\left(\left\vert x-x_{k}\right\vert\right)
\end{align}
where $N$ is the number of vortices, ${x}_{k}$ is the vortex positions
($\overline{x}_{k}$ denotes the complex conjugate of $x_{k}$),
$\Gamma_{i}$ is the vortex strengths, and $g_{\delta}(r)$ is the Gaussian
smoother with radius $\delta$,
\begin{align}
\label{eq:gaussian}
  g_{\delta}(r)=1-\exp\left(-\frac{r^{2}}{\delta^{2}}\right),
\end{align}
which is applied to avoid divergence as $x\rightarrow x_{k}$. The
propagation is carried out using the Euler forward scheme. At the
initial time, all vortices are located regularly in a long and
relatively thin rectangle, with the upper half of the vortices given
the opposite strength of the lower half (hence the sum of all
circulations is zero). This setup creates a velocity shear field,
which is an unstable configuration very similar to a Kelvin-Helmholtz
instability. The simulation starts with a homogeneous distribution and
evolves toward a more clustered distribution. This affects the number
of connections in the near field of the interaction lists since the
clustered distribution has a much larger variation in box-sizes.

\begin{figure}
  \includegraphics{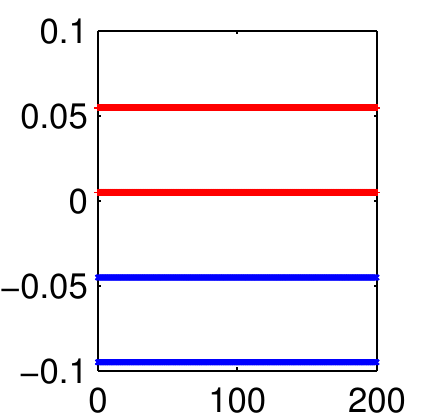}
  \includegraphics{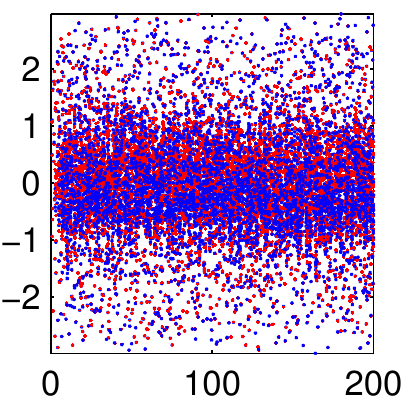}
  \caption{Distribution of points in vortex instability simulation at
    initialization and after 250 time-steps.}
  \label{fig:kh_pointplots}
\end{figure}

\subsubsection{Evaluating the autotuners}

We ran the vortex instability simulation with each of our autotuners,
as well as with a constant initial setting, to characterize the
effectiveness of our autotuning schemes.  Two problem sizes were used
to illustrate how the autotuners' efficiency varies according to
problem specifics.  For the large problem, the initial value of
$\nlevels$ was set to one less than optimal, which is a common
scenario for example when stepping up from a small prototype to a
full-scale production run. Table~\ref{tab:khinstability_tuning} shows
that for large problems, all autotuners are able to realize
significant performance improvements compared to an untuned run.  We
also see that AT3b performs slightly better than the other three
schemes. In the small run, constant factors dominate runtime and
tuning makes little difference. Here, AT3a makes costly tuning
attempts in the wrong direction and the performance is $2.5\%$ worse
than for the untuned case.  In the rest of the paper, we use AT3b
unless otherwise specified. This tuning scheme fulfills the
requirements specified in Section \ref{sec:dynamic_autotuning} better
than the other candidates.

\begin{table}[htdp]
  \begin{center}
    \begin{tabular}{lrrrrr}
     \hline
     \backslashbox{~~~~}{Scheme} & none & AT1 & AT2 & AT3a & AT3b \\
     \hline
     Small & 1 & 1.02 & 1.05 & 0.97 & 1.07 \\
     Large & 1 & 2.43 & 2.48 & 2.47  & 2.51 \\
     \hline
   \end{tabular}
  \end{center}
  \caption{Relative speedup of smaller ($N=16000$) and larger
    ($N=1.5$ million) vortex instability simulations using different
    tuning schemes.}
  \label{tab:khinstability_tuning}
\end{table}

\subsection{Rotating galaxy}

A two dimensional version of Newton's law of gravitation states that
the gravitational force felt by a particle $i$ from a particle $j$ of
mass $m_j$ is given by
\begin{align}
  \label{eq:Ugravity}
        F_{ij} = \frac{G m_j}{r_{ij}},
\end{align}
where $G$ is the gravitational constant and $r_{ij}$ is the distance
between the particles.
% +++ a nice discussion is found at
% http://physics.stackexchange.com/questions/30652/what-is-the-2d-gravity-potential
% (this is actually F = -grad V, with V a fundamental solution to the
% Laplace equation in 2D)

Equation \eqref{eq:Ugravity}, as written, contains a singularity as
$r_{ij} \to 0$, so a smoother is applied that prevents the velocity
for close particles from blowing up;
\begin{align}
  F_{ij} = \frac{G m_j}{\sqrt{\delta^{2}+r_{ij}^{2}}},
\end{align}
with $\delta$ a smoothing radius.

\begin{figure}
  \includegraphics{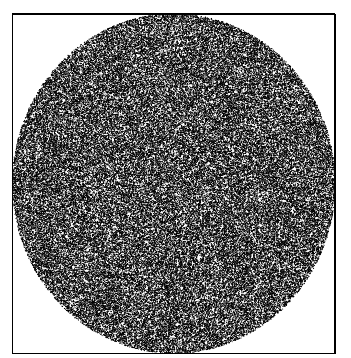}
  \includegraphics{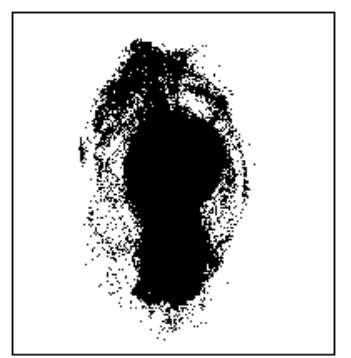}
  \caption{Distribution of points on disc in galaxy simulation at
    initialization and after 3000 time-steps.}
  \label{fig:galaxy_pointplots}
\end{figure}

In this simulation, $3\times10^5$ particles with equal mass are placed
uniformly in a disc and are given a velocity to start rotating as a
rigid body about the center of mass. The force acting on each particle
is calculated via the FMM with an error tolerance of $3 \times 10^{-8}$, and its velocity and position is updated
with the velocity St\"ormer-Verlet method
\cite[Chap.~3.1]{griebelMoldyn}. As the simulation evolves, the
distribution of particles within the disc becomes clustered and the
structure begins to resemble an elliptic galaxy, see
Figure~\ref{fig:galaxy_pointplots}.

\subsubsection{Initial tuning parameters}

\begin{figure}
  \includegraphics{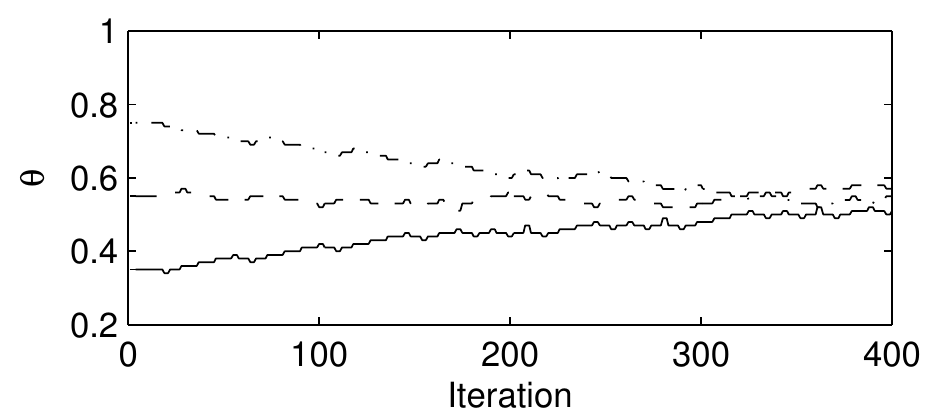}
  \caption{Evolution of $\theta$ for three different starting
    conditions.
    \label{fig:galaxy_startthets}}
\end{figure}

\begin{figure}
  \includegraphics{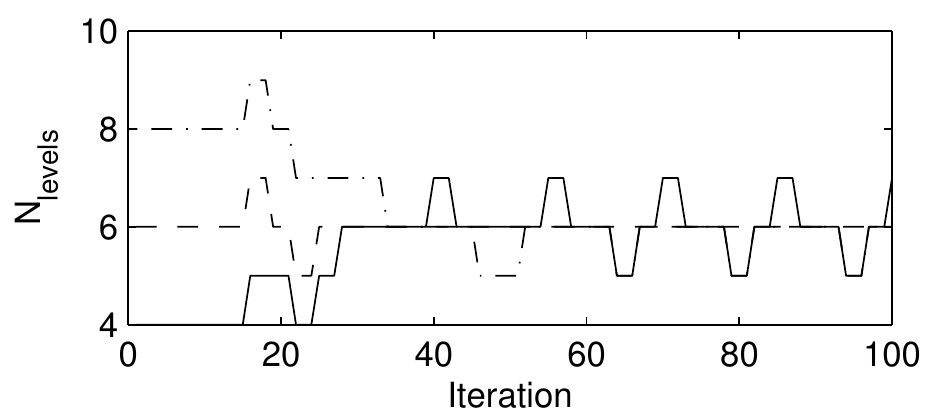}
  \caption{Evolution of $\nlevels$ for three different starting
    conditions. Note that the horizontal scale is shorter than in Figure
    \ref{fig:galaxy_startthets}.
    \label{fig:galaxy_startnlevels}}
\end{figure}

The importance of choosing initial tuning parameters is the most
apparent when running a short, relatively static simulation.  Our
galaxy simulation is therefore well-suited for a study of the initial
tuning if we simulate a small number of time-steps. In
Table~\ref{tab:galaxy_startconds}, we present the relative runtime of
the simulation with 300 time-steps for different sets of initial parameters. Because of
the coarse granularity and big marginal effect on program speed, the
optimal value of $\nlevels$ is found very quickly compared to the
optimal $\theta$ (see Figures \ref{fig:galaxy_startthets} and
\ref{fig:galaxy_startnlevels}).

Due to the small number of iterations and the large size of the
problem, this simulation is highly sensitive to bad initial tuning.
We can therefore consider these results to be a worst-case scenario
and expect that most simulations will experience a smaller impact due
to initial choices of parameters.

\begin{table}[htdp]
  \begin{center}
    \begin{tabular}{lrrrr}
     \hline
     \backslashbox{~~~~$\theta$, $p$}{$\nlevels$} & 4 & 5 & 6 & 7 \\
     \hline
     0.35, 14 & 1.16 & 1.05 & 1.02 & 1.03 \\
     0.55, 25 & 1.10 & 1.03 & 1\phantom{.00} & 1.04 \\
     0.75, 54 & 1.13 & 1.07 & 1.32 & 1.46 \\
     \hline
   \end{tabular}
  \end{center}
  \caption{The total runtime of a 300 time-step galaxy simulation with
    different initial tuning parameters. Time is normalized to the
    fastest total runtime. $p$ is the initial number of multipole
    coefficients. The relative error tolerance was set to $3 \times
    10^{-8}$.}
  \label{tab:galaxy_startconds}
\end{table}

\subsection{Impulsively started flow around a cylinder}

\begin{figure}
  \includegraphics[width=0.8\columnwidth,trim = 5cm 9cm 4cm
  8cm,clip]{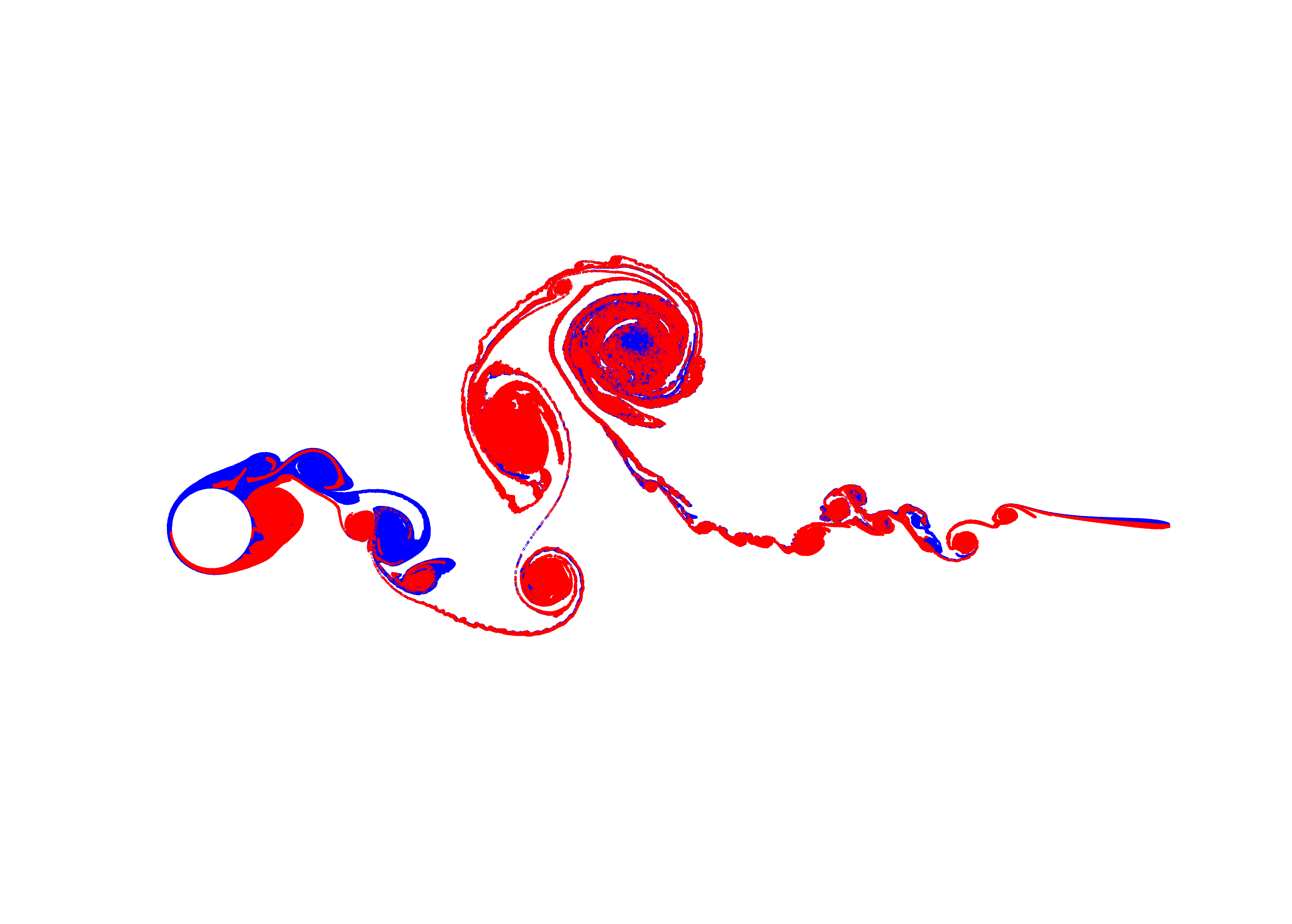}
  \caption{Sample vortex distribution for the rotating cylinder flow after 500 time steps.}
  \label{fig:cylinderflow}
\end{figure}

A vortex method was used to simulate the flow around a rotating
cylinder (peripheral speed one half the asymptotic flow velocity).
This method is based on the vorticity equation \cite{cottet08}, which
is obtained by calculating the rotation of Navier Stokes
Equations. For two dimensional incompressible flow, the equation can
be written as
\begin{align}
  \frac{\partial\omega}{\partial t}+
  \left(\vec{V}\cdot\nabla\right)\omega=\nu\nabla^{2}\omega.
\end{align}
This equation is discretized using point vortices with positions
$\vec{x}_{k}$ and strengths $\Gamma_{k}$, which moves freely with the
flow. The equation is separated into two steps, where in the
convection step, the particle positions are calculated according to
\begin{align}
  \frac{d\vec{x}_{k}}{dt} &= \vec{V}_{k}
\end{align}
and in the diffusion step, the vorticity is updated;
\begin{align}
  \frac{d\omega}{dt} &= \nu\nabla^{2}\omega.
\end{align}
The diffusion step is implemented using the vorticity redistribution
method (VRM) \cite{Shankar96}. The diffusion algorithm was also used
to merge close vortices as this can be done in the VRM by forcing the
circulation of a vortex to zero. This was performed every 10th step
and causes the regions with vorticity to have a quite homogeneous
distribution of vortices. The total distribution of vortices was
therefore characterized by regions with either a homogeneous vortex
distribution, or no vortices at all, see
Figure~\ref{fig:cylinderflow}. The no slip boundary conditions was
solved by adding a line of vortices at a distance of
$\sqrt{0.5\nu\Delta t}$ from the boundary and adapting the strength of
these vortices to give zero tangential flow on the boundary. This
method is similar to the suggested solution by Chorin \cite{Chorin73}
and is how vorticity is introduced into the flow.

The flow velocity is calculated using the method of images, which
gives an analytical solution to the continuity equation for a cylinder
flow \cite{milne68}. Using complex numbers, the velocity $V$ at
position $z$ is obtained as
\begin{align}
  \label{eq:velocity}
  V &= V_{\infty}\left(1-\frac{R^{2}}{x^{2}}\right)+
  \frac{1}{2\pi i}\sum_{k=1}^{N}\Gamma_{i}
  \left(\frac{g_{\delta}\left(\left\vert \overline{x}-
    \overline{x}_{k}\right\vert\right)}{\overline{x}-\overline{x}_{k}}-
  \frac{g_{\delta}\left(\left\vert \overline{x}-
    \frac{R^{2}}{x_{k}}\right\vert\right)}{\overline{x}-
    \frac{R^{2}}{x_{k}}}\right),
\end{align}
where $g_{\delta}(r)$ is the Gaussian smoother in \eqref{eq:gaussian}.
This expression includes the velocity contribution from all vortices
at positions $x_{k}$ (again, $\overline{x}_{k}$ in \eqref{eq:velocity}
is the complex conjugate of $x_{k}$).

All interactions between the vortices are evaluated with the FMM.  Due
to the mirror vortices inside the cylinder, the amount of vortices is
twice as many as the number of evaluation points. The mirror vortices
are very densely packed inside the cylinder, and especially so close
to the cylinder center. Although our particular implementation only
applies for a cylinder, by using conformal mappings the technique with
mirror vortices can be used to simulate other geometries as well
\cite{Deglaire08,vertaxis2D}.

The convection step is carried out using the fourth order Runge-Kutta
method, while the diffusion step is handled in a single step using the
$O(\Delta t)$ method described in \cite{Shankar96}. The higher order
for the convection (compared to diffusion) is motivated by the fact
that the problem is convection dominated for high Reynolds number and
that increasing the VRM order is computationally very expensive.

The simulations were performed for an impulsively started flow,
meaning that there were no vortices at the first time-step. The
Reynolds number for the simulation was chosen to be 10000.

Since both the distribution geometry as well the number of particles
varies greatly during this simulation, it can be considered a stress
test of our implementation and the auto-tuning schemes.

\subsubsection{Capping the tuning cost}

\begin{figure}
  \includegraphics{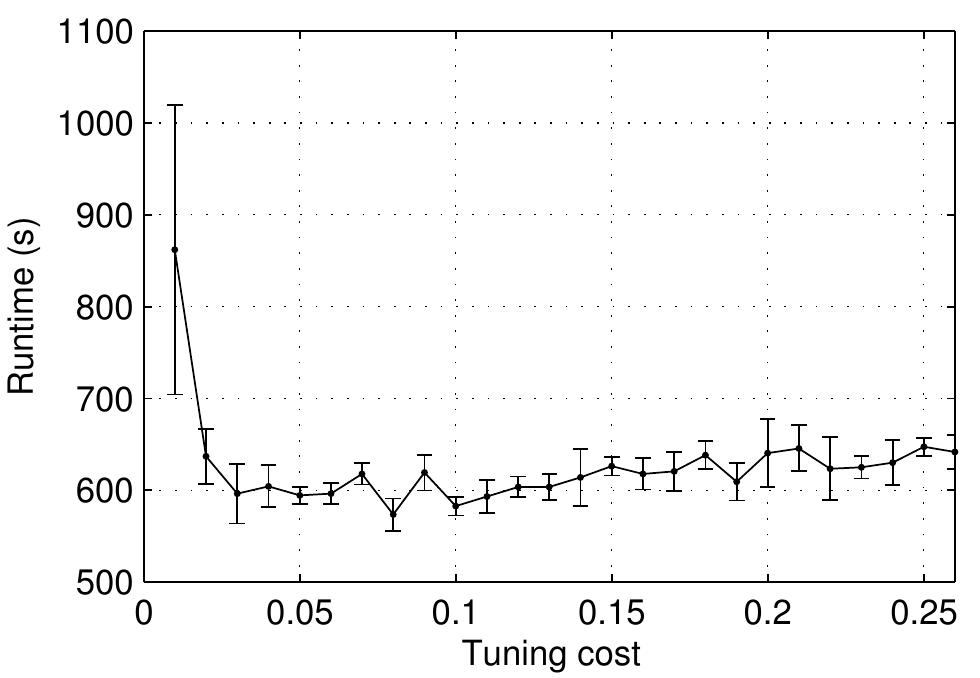}
  \caption{Runtime of cylinder flow simulation for 400 time-steps
    using AT3b with varying maximum tuning cost ($\texttt{cap}$).  The
    simulation was repeated 5 times for each tuning cost, and the
    error bars indicate one standard deviation. }
  \label{fig:cylinderflow_levcost_runtime}
\end{figure}

\begin{figure}
  \includegraphics{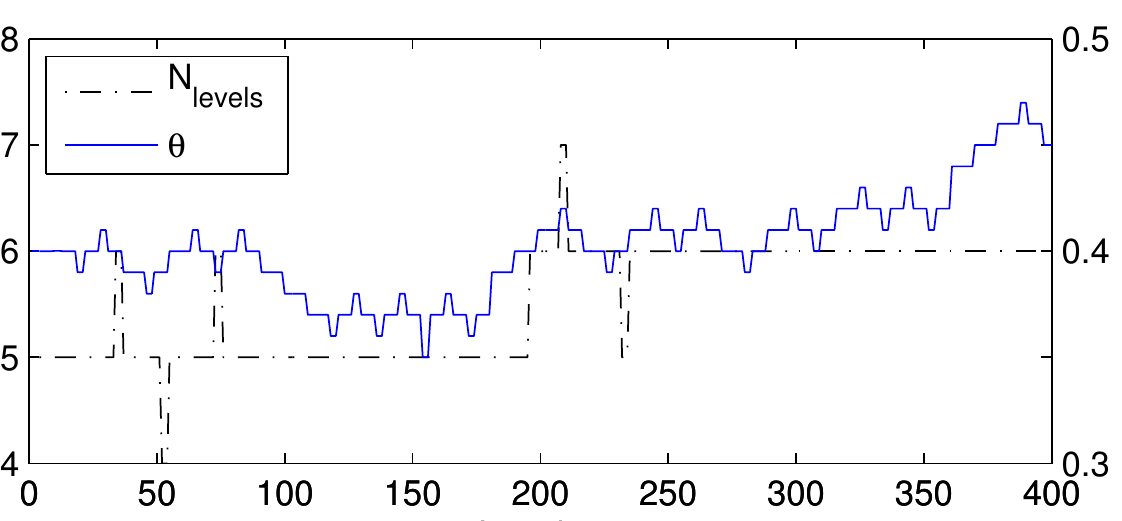}
  \includegraphics{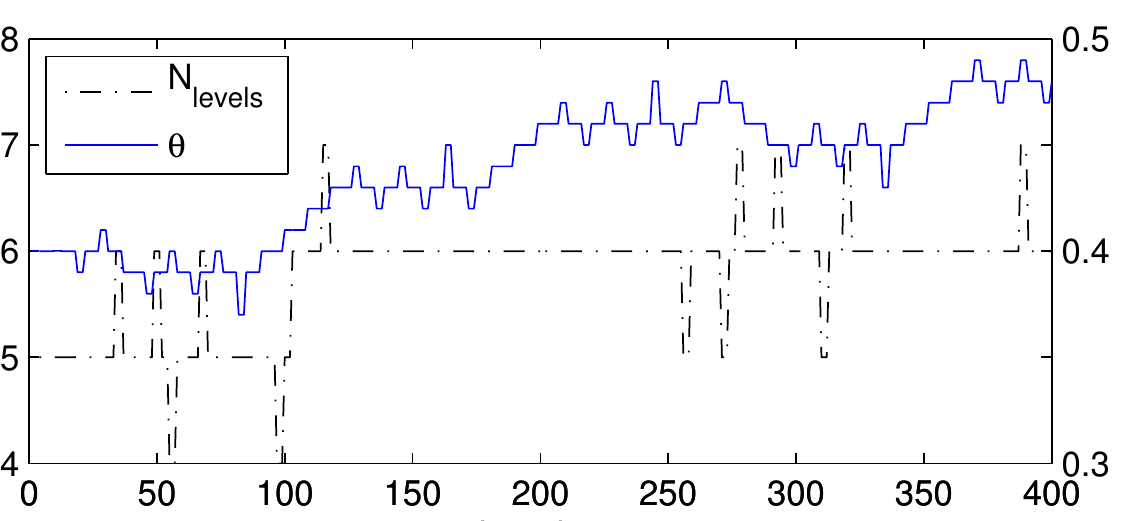}
  \caption{Timeseries showing evolution of $\nlevels$ and $\theta$
    over the course of a cylinder flow simulation run. \emph{Top:} Max
    tuning cost $\texttt{cap}=4\%$, slightly too low to capture the
    optimal switching points. \emph{Bottom:} $\texttt{cap}=12\%$,
    slightly higher than necessary. \label{fig:cap_timeseries}}
\end{figure}

AT3b allows the user to arbitrarily select $\texttt{cap}$, the maximum
expected cost of tuning $\nlevels$. If $\texttt{cap}=0$, no tuning of
$\nlevels$ is done, while if $\texttt{cap}$ is large enough,
$\nlevels$ is adjusted every time-step. In Figure
\ref{fig:cylinderflow_levcost_runtime}, we have repeated the flow
simulation with varying $\texttt{cap}$. The proportional rise in
simulation runtime after about $\texttt{cap}=0.1$ suggests that, even
for a rapidly evolving simulation such as this, tuning need not cost
more than 10\% for it to accurately capture switching times in
$\nlevels$.  For simulations that evolve more slowly, $\texttt{cap}$
can be set lower, between 5--10\%. To give a feeling for how
$\texttt{cap}$ affects the tuner's behavior, Figure
\ref{fig:cap_timeseries} shows the value of $\nlevels$ and $\theta$
over time with $\texttt{cap}=4\%$ and $12\%$.

%**************************************************************************

\section{Conclusions}
\label{sec:conclusions}

In this paper we have presented a variant of the FMM and shown that it
can be efficiently adapted for parallelization on hybrid computer
systems, exploiting both multiple threads and an accelerator. We base
the parallel FMM scheme on the heterogeneity in the algorithm, and by
using a dynamic autotuning technique our implementation can exploit
systems with different hardware characteristics and adapt to different
problem settings without the need for explicitly modifying computer
architecture- or problem-dependent parameters.

% (duplicate from the intro)
%% We argue that the approach of basing implementations on
%% heterogeneous computer systems on a hybrid approach using th
%% inherent heterogeneity of \emph{the algorithm} can be very fruitful
%% for many other computational science kernels, apart from the FMM.

% +++
% While it is difficult to determine the programming cost exactly, our
% approach to using the GPU requires the least possible effort. With
% the algorithmic modifications described in Section
% \ref{sec:hybridfmm}, the P2P step is trivially the simplest to
% implement on a GPU, and the rest of the FMM algorithm is easily
% parallelizable on the CPU. Executing any other combination of steps
% on the GPU would lead to much higher code complexity, so we are
% pleased to see that our approach yields a good performance
% improvement over the CPU-only code.

We also come to several other conclusions that we believe are of
fairly general character. Firstly, optimal performance can generally
not be expected when a heterogeneous computer is load balanced. In
fact, we have shown that the degree of load \emph{im}balance can
\emph{at best} only be used as an indicator of where to look for
well-tuned and efficient core usage.

Secondly, a robust autotuning strategy which does not make strong
complexity assumptions on the algorithm at hand will require a certain
degree of heuristics. To some extent this observation is connected to
the previous point since, if we cannot completely rely on the
dynamically measured degree of load balance, then clearly some type of
trial-and-error has to be employed.

Thirdly, if some design variables affect the performance more
drastically than others, a performance budget strategy of some kind
should be implemented. In the implementation reported here this was
the case for the variable $\nlevels$ where the cost of varying this
variable can be measured on the fly so as to avoid too frequent
changes.

% +++
% Since cost estimation of $\nlevels$ turned out to be useful, one
% naturally considers introducing a similar mechanism for tuning
% $\theta$. One approach would be to replace the W-cycle idea with a
% schedule for the entire $\theta$ range.  However, we haven't been
% able to show a need for even the W-cycle in real simulations, so
% going a step further is doubly unnecessary. The current cost of
% tuning $\theta$ is usually far below reasonable thresholds.

The utility of GPU-based accelerators is still a much-debated question
in the scientific computing community. While an increasing proportion
of large-scale computing machinery include GPUs, smaller
university-level computers are still predominantly pure CPU machines,
and typical desktop computers have GPUs that are more suited to media
rendering than scientific applications. Our contribution to this
debate is the transparent usage of one GPU, with a performance
improvement that matches (or even exceeds) the performance one would
expect from an expensive CPU upgrade. Table~\ref{tab:cpugpu} gives the
relative performance of our code with or without the GPU for each of
our experiments. We see that the 4x speedup in the static experiment
forms a reasonable upper limit of hybrid acceleration, and as expected
this limit is only realized for sufficiently large problems.

\begin{table}[htdp]
  \begin{center}
    \begin{tabular}{lrrr}
      \quad & Vortex instability & Galaxy & Cylinder flow \\
      \hline
      CPU (s) & 425 & 1094 & 1451 \\
      CPU+GPU (s) & 379 & 324 & 367 \\
      \hline
      Speedup & 1.12 & 3.37 & 3.95 \\
    \end{tabular}
  \end{center}
  \caption{Performance comparison of CPU-only and hybrid code for the
    experiments conducted in the paper.}
  \label{tab:cpugpu}
\end{table}

\subsection{Reproducibility}
\label{subsec:reproduce}

The FMM implementation described in this paper is available for
download via the second author's
web-page\footnote{\url{http://user.it.uu.se/~stefane/freeware}}. The
code compiles both in a CPU-only version and in a hybrid version and
comes with a convenient Matlab mex-interface. The hybrid version
requires Cuda and an Nvidia GPU to function. Along with the code,
automatic Matlab-scripts that repeat the numerical experiments
presented here are also distributed.

%**************************************************************************

\section*{Acknowledgment}

This work was financially supported by the Swedish Research Council
within the UPMARC Linnaeus center of Excellence (M.~Holm, S.~Engblom,
S.~Holmgren) and the Swedish Energy Agency, Statkraft AS, Vinnova and
Uppsala University within the Swedish Centre for Renewable Electric
Energy Conversion (A.~Goude).

%**************************************************************************

\newcommand{\doi}[1]{\href{http://dx.doi.org/#1}{doi:#1}}
\newcommand{\available}[1]{Available at \url{#1}}
\newcommand{\availablet}[2]{Available at \href{#1}{#2}}

\bibliographystyle{abbrvnat}
\bibliography{eghh}

\end{document}